\input{aipcheck}

\documentclass[
   ,draft            % use draft while you are working on the paper
  ]
  {aipproc}

\layoutstyle{8x11single}

\begin{document}

\title{Measurement of the $\nu_\mu$ Charged Current $\pi^+$  Production  to Quasi-elastic Scattering Cross Section }


\author{Jaroslaw A. Nowak, for the MiniBooNE Collaboration}{
  address={Department of Physics and Astronomy \\
  Louisiana State University\\ Baton Rouge, LA 70803}
}

\begin{abstract}
Using high statistics samples of charged current interactions, MiniBooNE reports a model independent measurement of the single charged pion production to quasi-elastic cross section ratio on mineral oil without corrections for pion re-interactions in the target nucleus \cite{AguilarArevalo:2009eb}. The result is provided as a function of neutrino energy in the range 0.4 GeV < E < 2.4 GeV with 11\% precision in the region of highest statistics. 
\end{abstract}

\maketitle

%%%%%%%%%%%%%%%%%%%%%%%%%%%%%%%%%%%%%%%%%%%%
%% MAINMATTER
%%%%%%%%%%%%%%%%%%%%%%%%%%%%%%%%%%%%%%%%%%%%

\section{Introduction}

Future neutrino experiments will operate in the few-GeV neutrino energy range, where quasielastic scattering (CCQE) and pion production channels are dominant interactions. The charged current pion production interactions  account for the biggest background for the neutrino oscillation disappearance measurements.  Recently, three experiments (K2K \cite{k2k1}, MiniBooNE \cite{miniboone1, miniboone2} and SciBooNE \cite{sciboone1, sciboone2}) have presented results for CC pion production cross section in the few-GeV neutrino energy region on nuclear targets. With high statistic samples it is possible to determine neutrino cross sections in this region with better precision than data from previous experiments\cite{Radecky:1981fn, Kitagaki:1986ct,K2K:2008eaa} .  Here, we present the ratio of  cross sections of the charged current single pion production ($CC1\pi^+,\  \nu_\mu X \to \mu^- \pi^+ X'$) and quasielastic scattering ($CCQE, \ \nu_\mu n \to \mu^- p$) as obtained from the MiniBooNE experiment. In this  measurement the largest sources of uncertainty coming  from the neutrino  flux determination largely cancels out.

The Booster Neutrino Beam at Fermilab provides a neutrino source which is particularly well-suited to make this measurement; about 40\% of $\nu_\mu$ neutrino interactions in MiniBooNE are expected to be CCQE and 24\%  $CC\pi^+$. The beam itself is composed of 93.6\% $\nu_\mu$ with a mean energy of about 800 MeV and 5.9\% (0.5\%)  $\bar \nu_\mu$  ($\nu_e$) contamination \cite{AguilarArevalo:2008yp}.

The neutrinos are detected in the MiniBooNE detector \cite{AguilarArevalo:2008qa}, a 12.2 m diameter spherical tank  filled with 818 tons of undoped mineral oil located 541 m downstream of the beryllium target. At the energies relevant to this analysis, the products of the interactions produce primarily Cerenkov   light with a small fraction of scintillation light \cite{AguilarArevalo:2008qa}. The light is detected by 1280 8-inch photomultiplier tubes (PMTs) which line the MiniBooNE inner tank. This inner tank region is optically isolated from a surrounding veto region, instrumented with 240 PMTs, that serves to reject incoming cosmic rays and partially contained neutrino interactions.

Neutrino interactions within the detector are simulated with the v3 NUANCE event generator \cite{Casper:2002sd}. CCQE interactions on carbon are generated using the relativistic Fermi gas model \cite{Smith:1972xh} tuned to better describe the observed distribution of $CCQE$ interactions in MiniBooNE \cite{:2007ru}. Resonant $CC1\pi^+$ events are simulated using the Rein and Sehgal (R-S) model \cite{Rein:1980wg}, as implemented in NUANCE with an axial mass $M^{1\pi}_A = 1.1~GeV$. The angular distribution of the decaying pions in the center of mass of the recoiling resonance follows the helicity amplitudes of \cite{Rein:1982pf}. In MiniBooNE, $87\%$ of $CC1\pi^+$ production is predicted to occur via the $\Delta (1232)$ resonance, but 17 higher mass resonances and their interferences, as well as a nonresonant background  that accounts for roughly 6\% of $CC1\pi^+$ events, are also included in the model. Coherently produced $CC1\pi^+$ events are generated using the R-S model \cite{Rein:1982pf} with the R-S absorptive factor replaced by NUANCE's pion absorption model and the overall cross section rescaled to reproduce MiniBooNE's recent measurement of neutral current coherent $\pi^0$ production \cite{AguilarArevalo:2008xs}. Coherent $\pi^+$ production is predicted to compose less than 6\% of the MiniBooNE $CC1\pi^+$ sample due to the small coherent cross section \cite{Hasegawa:2005td, Hiraide:2008eu}  and the dominance of the $\Delta^{++}$ resonance. A GEANT3-based detector model  simulates the response of the detector to particles produced in these neutrino interactions.

\section{Analysis}

The $CCQE$ and $CC1\pi^+$ reconstruction requires a detailed model of light production and propagation in the tank to predict the charge distribution for a given vertex and muon angle. The muon vertex, track angle, and energy, are found with a maximal likelihood fit, with the energy being determined from the total tank charge \cite{Patterson:2009ki} . In the case of $CC1\pi^+$ events, only one track is reconstructed and assumed to be a $\mu^-$.  The neutrino energy for both samples is reconstructed from the observed muon kinematics, treating the interaction as a 2-body collision and assuming that the target nucleon is at rest inside the nucleus:

\begin{equation}
E_\nu = \frac{1}{2} \frac{2m_p E_\mu + m_1^2 - m_p^2 -m_\mu^2}{m_p - E_\mu \cos \theta \sqrt{E_\mu^2 - m_\mu^2}}
\end{equation}

Here $m_p$ is the mass of the proton, $m_\mu$ is the mass of the muon, $m_1$ is the mass of the neutron in $CCQE$ events and of the $\Delta(1232)$ in $CC1\pi^+$, $\theta_\mu$ is the reconstructed angle of the muon with respect to the beam axis (in the lab frame), and $E_\mu$ is the reconstructed muon energy.

The cross-section for a neutrino interaction is defined as:

\begin{equation}
\sigma_X =\frac{N_X}{\Phi_\nu N_{targets}}
\end{equation}

where N is the true number of events of type X($CC1\pi^+$ or $CCQE$),  $N_{targets}$ is the number of nuclear targets in the detector, and $\Phi$ is the flux of incident neutrinos.
Because the incident neutrino flux and the number of nuclear targets are fixed properties of the beam and the detector (i.e. independent of the process we are considering), $N_{targets}$  and $\Phi_\nu$ must be the same for both processes. Therefore we can write

\begin{equation}
\sigma_{ccpip} (E_\nu) =\frac{N_{ccpip} (E_\nu) }{ N_{ccqe} (E_\nu)   }  \sigma_{ccqe}(E_\nu). 
\end{equation}

With all the correction terms put together, the cross section ratio in each energy bin $i$  is:
\begin{equation}
\frac{\sigma_{1\pi^+,i}}{\sigma_{QE,i}} = \frac{\epsilon_{QE,i} \ast \sum_j U_{1\pi^+,ij } \ast f_{1\pi^+,j} \ast N_{1\pi^+,j}^{cuts} } {  \epsilon_{1\pi^+,i} \ast \sum_j U_{QE,ij } \ast f_{QE,,j} \ast N_{QE,,j}^{cuts} }
\end{equation}

where the subscript $i$ runs over bins in true neutrino energy, subscript $j$ indexes bins in reconstructed neutrino energy, $N_{X-cuts}$ denotes the number of events passing cuts
for $X = CC1\pi^+$, $CCQE$, $f$ denotes the signal fraction, $\varepsilon$ denotes the cut efficiency, and $U$ is a neutrino energy unsmearing matrix that acts on a reconstructed distribution to return the true distribution.

To map reconstructed to true energy, we form a migration matrix $U_{ij}$ representing the number of MC events in bin $i$ of reconstructed energy and bin $j$ of true energy. We then normalize each reconstructed energy bin to unity to obtain an unsmearing matrix. This is equivalent to a Bayesian approach discussed in \cite{D'Agostini:1994zf}; it differs from the standard matrix inversion method in that the resulting unsmearing matrix is biased by the MC distribution used to generate it. We account for this in our uncertainties by including a variation in the MC distribution used to generate the matrix. Because we have good data/MC agreement, this effect is small. The advantage of this method is that it avoids the problems of numerical instability and the magnification of statistical errors which occur in matrix inversion. This unsmearing procedure also proved insensitive to variations in neutrino energy reconstruction, confirming that it performs as intended.

\section{Results}

For our primary measurement, we define $CC1\pi^+$  signal as events with exactly one $\mu^-$ and one $\pi^+$ escaping the struck nucleus (which we call $CC1\pi^+$-like events) and
CCQE signal as those with exactly one $\mu^-$ and no pions ($CCQE$-like events). Both event classes may include any number of protons or neutrons, but no other types of hadrons.
The observed cross section ratio is then defined as the ratio of $CC1\pi^+$-like to $CCQE$-like events and thus has not been corrected for re-interactions in the struck nucleus.

\begin{figure}
\begin{tabular}{cc}
  \includegraphics[height=.4\textheight]{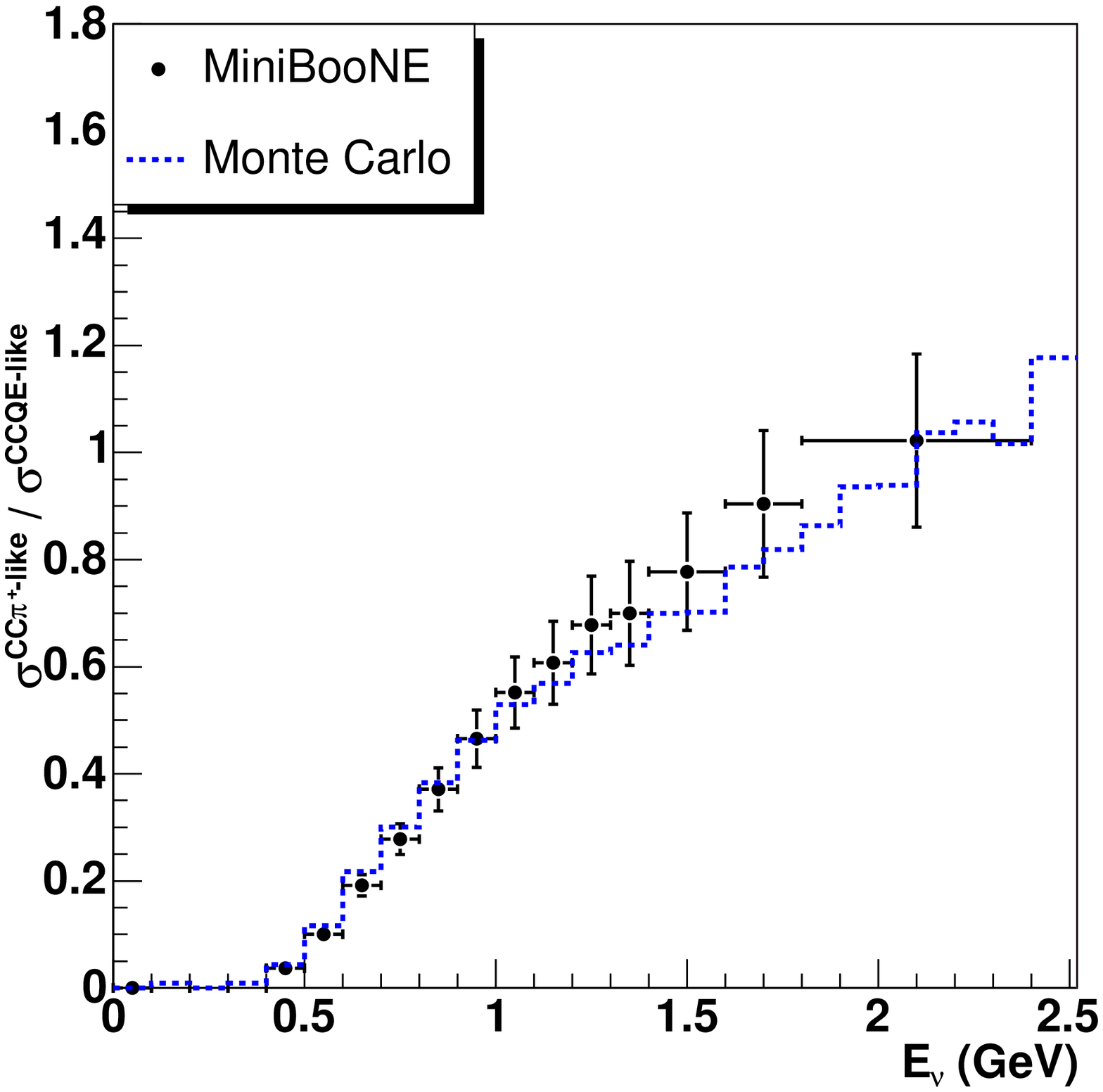}
&
  \includegraphics[height=.4\textheight]{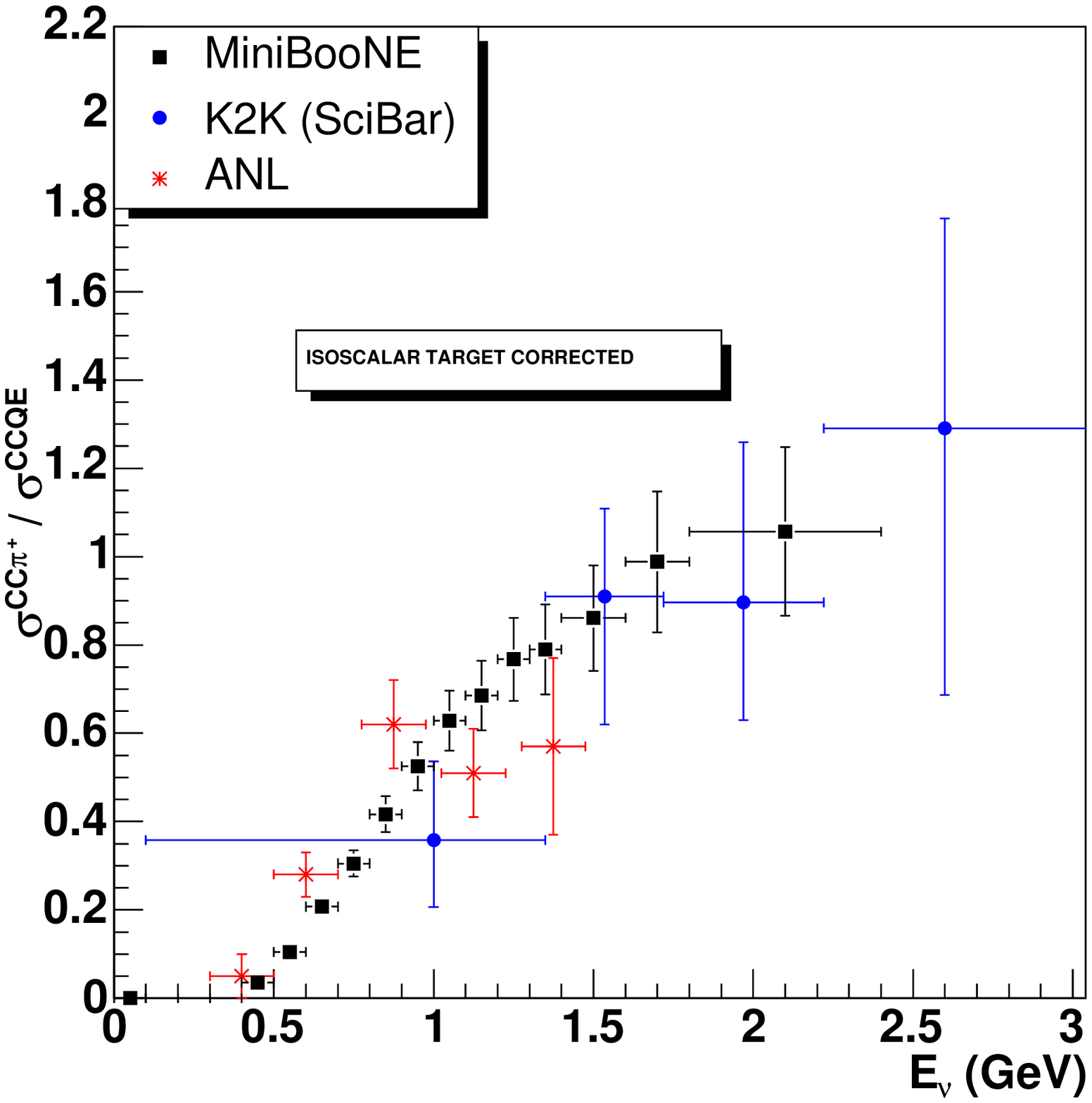}
\end{tabular}
\caption{Left: Observed $CC1\pi^+$-like/$CCQE$-like cross section ratio on $CH_2$, including both statistical and systematic uncertainties, compared with the MC prediction \cite{Casper:2002sd}.  The data have not been corrected for hadronic re-interactions.Right:  FSI-corrected $CC1\pi^+$ to $CCQE$ cross section ratio on $CH_2$ compared with results from ANL (D2) \cite{Radecky:1981fn} and K2K ($C_8H_8$) \cite{K2K:2008eaa}. The data have been corrected for final state interactions and re-scaled for an isoscalar target .\label{fig}}
\end{figure}

Figure \ref{fig} shows the observed $CC1\pi^+$-like to $CCQE$-like ratio extracted from the MiniBooNE data, including statistical and systematic uncertainties.

The systematic uncertainties on the cross section ratio arise from five main sources: the neutrino flux (which largely cancels in the ratio), the neutrino interaction cross sections (which affect the background predictions), the target nucleon momentum distribution (which accounts for the model dependence of our unfolded neutrino energy), hadron re-interactions in the detector, and the detector simulation (which describes light propagation in the oil). In the region of highest statistics (about 1 GeV), there is roughly an 8\% fractional error on the ratio resulting from hadron re-scattering in the detector, 6\% from neutrino cross sections, 4\% from the detector simulation, 2\% from the nucleon momentum distribution, 2\% from the neutrino flux, and 2\% from the statistics of the two samples.

The cross section ratio reported by all prior experimental measurements \cite{Radecky:1981fn, Kitagaki:1986ct, K2K:2008eaa}  has been one in which the effects of  final state interactions (FSI) in the target nucleus have been removed using MC or deuterium was used as a target. Solely for the purpose of comparison, we now extract a similarly corrected value. The FSI-corrected ratio is defined as the ratio of $CC1\pi^+$ to $CCQE$ events at the initial vertex and before any hadronic re-interactions. Thus, the signal fractions and cut efficiencies for the FSI-corrected ratio include corrections for intra-nuclear hadron re-scattering based on the MC's model for nuclear effects  \cite{Casper:2002sd}.  The measurement proceeds exactly as for the observed ratio, except that now we define $CC1\pi^+$ and $CCQE$, rather than $CC1\pi^+$-like and $CCQE$-like, events as signal for the respective samples.

Here we limit our comparison to those experiments which reported both $CCQE$ and $CC1\pi^+$ cross sections, using the same energy bins for each of these interactions, so as to facilitate comparison with our measured $CC1\pi^+/CCQE$ ratio. Our result agrees with both ANL, which used a deuterium target, and K2K, which used $C_8H_8$. In order to make this comparison, the
MiniBooNE and K2K results have been re-scaled to an isoscalar target. To perform this correction, we rescale the ratio by a factor of  $(1 - r) s_p$, where $r$ is the ratio of
neutrons to protons in the target and $s_p$ is the fraction of $\pi^+$ production that is predicted (by MC) to occur on protons. The resulting scaling factor is 0.80 for MiniBooNE;
for K2K we use the factor of 0.89 provided in \cite{K2K:2008eaa}. The results have not been corrected for their differing nuclear targets nor for the application of explicit invariant mass
requirements (although the latter are similar). ANL used an explicit cut on invariant mass W < 1.4 GeV \cite{Radecky:1981fn} . While no invariant mass cut is used in this analysis, the MiniBooNE spectrum is such that $CC1\pi^+$ events occur only in the region $W < 1.6~GeV$; similarly, K2K's measurement covers the region W < 2 GeV \cite{K2K:2008eaa} .

\begin{table}
\begin{tabular}{l|r|c}
\hline
      \tablehead{1}{c}{b}{$E_\nu$ \\ (GeV) }
  & \tablehead{1}{c}{b}{$CC1\pi^+ / CCQE$ (FSI corrected)}
  & \tablehead{1}{c}{b}{$CC1\pi^+ -like/ CCQE-like$ (observed)}   \\
\hline 

$0.45\pm0.05$ & $0.045\pm0.008$  &  $0.036\pm0.005$\\ 
$0.55\pm0.05$ & $0.130\pm0.018$  &  $0.100\pm0.011$\\
$0.65\pm0.05$ & $0.258\pm0.033$   & $0.191\pm0.019$\\
$0.75\pm0.05$ & $0.381\pm0.047$   & $0.278\pm0.028$\\
$0.85\pm0.05$ & $0.520\pm0.064$   & $0.371\pm0.040$\\
$0.95\pm0.05$ & $0.656\pm0.082$   & $0.465\pm0.053$\\
$1.05\pm0.05$ & $0.784\pm0.100$   & $0.551\pm0.066$\\
$1.15\pm0.05$ & $0.855\pm0.114$   & $0.607\pm0.077$\\
$1.25\pm0.05$ & $0.957\pm0.132$   & $0.677\pm0.091$\\
$1.35\pm0.05$ & $0.985\pm0.141$   & $0.700\pm0.097$\\
$1.5  \pm0.1   $ & $1.073\pm0.157$   & $0.777\pm0.109$\\
$1.7  \pm0.1   $ & $1.233\pm0.207$   & $0.904\pm0.137$\\
$2.1  \pm0.3   $ & $1.318\pm0.247$   & $1.022\pm0.161$\\
\hline
\end{tabular}
\caption{The MiniBooNE measured $CC1\pi^+$ to $CCQE$ (as in Figure 1 but without the isoscalar correction) and $CC1\pi^+$-like to $CCQE$-like (Figure 1) cross section ratios on $CH_2$ including
all sources of statistical and systematic uncertainty}
\label{tab:a}
\end{table}

The dominant reason for the difference between the ratios presented in Figure 1 is intra-nuclear pion absorption in $CC1\pi^+$ events, which cause these events to
look CCQE-like. As a result of $\pi^+$ absorption, a significant number of $CC1\pi^+$ events migrate from $CC\pi^+$ sample to $CCQE$-like one. Thus, the FSI-corrected ratio,  is 15\% to 30\% higher than the observed ratio in our energy range.

In summary, MiniBooNE has measured the ratio of $CC1\pi^+$-like to $CCQE$-like events for neutrinos with energy $0.4 GeV < E_\nu < 2.4 GeV$ incident on $CH_2$. This is the first time such a ratio has been reported. Additionally, the ratio of the $CC1\pi^+$ and $CCQE$ cross sections at the vertex has been extracted using MC to remove the effects of final state interactions, in order to facilitate comparison with previous experimental measurements. The results are summarized in Table I. The measured ratios agree with prediction  \cite{Casper:2002sd, Smith:1972xh, Rein:1980wg}  and previous data  \cite{Radecky:1981fn, Kitagaki:1986ct, K2K:2008eaa}.

\end{document}